\documentclass[aip,jap,
amssymb
]{revtex4-1}
\usepackage{graphicx}

\begin{document}

\title{Optimised low-loss multilayers for imaging with sub-wavelength resolution in the visible wavelength range}

\author{Anna Pastuszczak}
\email[Electronic mail: ]{anna.pastuszczak@igf.fuw.edu.pl}

\author{Rafa{\l} Koty{\'n}ski}
\affiliation{Faculty of Physics, University of Warsaw, Pasteura 7, 02-093 Warsaw, Poland}

\begin{abstract}
We optimise the effective skin-depth and resolution of Ag--TiO$_2$, Ag--SrTiO$_3$, and Ag--GaP multilayers for imaging with sub-wavelength resolution.
In terms of transmission and resolution the optimised multilayers outperform simple designs based on combined use of effective medium theory, impedance matching and Fabry-Perot resonances. For instance, an optimised Ag--GaP multilayer consisting of only $17$ layers, operating at the wavelength of $490$~nm and having a total thickness equal to one wavelength, combines $78\%$ intensity transmission  with a resolution of $60$~nm.
 It is also shown that  use of the effective medium theory leads to sub-optimal multilayer designs with respect to the trade-off between the skin depth and resolution already when the period of the structure is on the order of $40$~nm or larger.
\end{abstract}


\maketitle

\section{Introduction}
The presence of losses is a major limiting factor in the development of practical plasmonic devices. Here, we focus our interest on metal-dielectric multilayers for guiding optical wavefronts  consisting of both propagating and evanescent planewaves which may carry information about objects with sub-wavelength dimensions~\cite{anantharamakrishna2003raa,wood2006,belov2006prb,Podolskiy:apl2007,Scalora2007opex,Li:jap2007,Wang:Optexpr2008,vincentipra2009,Ahmadi:jap2009,kotynski_stefaniuk2009,Kotynski:oer2010,pendry:njp2010, Zapata:ao2010,Boltasseva2010}. The mechanism of such optical guidance may be attributed to coupling between surface plasmon polariton modes existing at the metal-dielectric interfaces, to negative refraction in photonic crystals, to effective optical anisotropy of the multilayer, to Fabry-Perot resonances, and typically to the interplay of these phenomena.
In effect, an appropriately designed multilayer is  capable of projecting the optical wavefront in between its external boundaries almost without diffraction including sub-wavelength details. Unfortunately, the permittivity of metals is complex-valued in the visible wavelength range resulting in a limited penetration depth inside the structure.
For instance the skin depth of bulk silver is   smaller than $20$~nm in the range of visible wavelengths.

There exist several ways of mitigating losses such as splitting the single-layer perfect lens~\cite{pendry2000nrm,fang2005sdl,melville2005sri} into a multilayer with thin metallic layers and compensating optical losses with gain,~\cite{anantharamakrishna2003raa,pendry:njp2010} using a high index dielectric with silver, which shifts the operational wavelength towards the red,~\cite{Stockman:NanoLett2005} or using semiconductors instead of silver for the wavelength ranges in between far-UV to far-IR.~\cite{Taubner:science2006, Hoffman:natmat2007,Vincenti:jap2009} Gain-tunable superresolution was recently considered along with the effect of gain on formation of optical vortices at the layer boundaries.~\cite{vincentipra2009} Reflection and transmission coefficients of the multilayer strongly depend on the termination conditions and in particular a symmetrical coating of the multilayer with a dielectric layer on both external boundaries leads to an increased transmission.~\cite{Scalora2007opex}
Simple yet successful designs of metal-dielectric superlenses for imaging with sub-diffraction resolution were designed with  the effective medium theory (EMT)~\cite{wood2006,belov2006prb} and recently refined beyond the second order Taylor expansion.~\cite{Podolskiy:apl2007} The canalization regime of transmission~\cite{belov2006prb} assumes that three effective permittivity and overall thickness  conditions  are fulfilled at the same time -the effective transverse permittivity of the multilayer is equal to that of the host medium $\varepsilon_x=1$, the axial permittivity is equal to  $\varepsilon_z^{-1}=0$, and the total thickness of the multilayer satisfies the Fabry-Perot condition for transmission. However, the role of the impedance matching condition was later questioned.~\cite{Li2007} In fact, multilayers with effective transverse permittivity $\varepsilon_x=0$ alone or the axial permittivity $\varepsilon_z^{-1}=0$ alone were also proposed,~\cite{Wang:Optexpr2008} with larger absolute permittivity of metal than the absolute permittivity of dielectric in the first case  and an opposite situation in the latter. Another approach assumes the use of complementary anisotropic slabs supporting negative refraction, where each slab consists of a silver-dielectric multilayer.~\cite{Li:jap2007} Moreover, engineering of the point spread function of the multilayer with sub-wavelength full-width-at-half-maximum ($FWHM$) may be conveniently cast into the framework of Fourier optics adjusted to include the evanescent waves.~\cite{Ahmadi:jap2009,kotynski_stefaniuk2009,Kotynski:oer2010,Zapata:ao2010} Finally, numerical optimisation was used to improve $FWHM$ in an impedance matched and impedance mismatched multilayer.~\cite{Li_josaa_2009}

\begin{figure}
 \begin{center}
\includegraphics{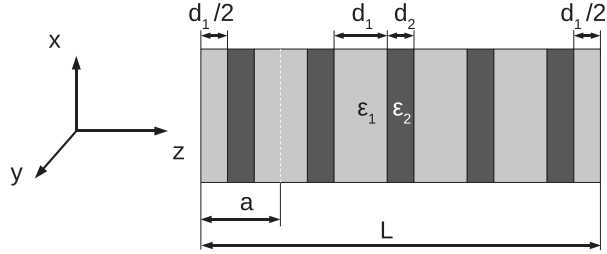}
\end{center}
\caption{The geometry of a silver-dielectric multilayer. An elementary cell of the structure $a$ consists of a silver layer with a thickness $d_2$ and two dielectric coating layers, each with a thickness $d_1 / 2$.}
\label{fig_layers}
\end{figure}

\section{Background}
\label{intro}

The dispersion relation of a periodic multilayer for a TM-polarised monochromatic planewave takes the form~\cite{Li2007, Wu2003}
\begin{equation}
cos(k_{B} \,a) = cos(k_{1} d_1) cos(k_{2} d_2) \\
- \frac{1}{2} \left(\frac{k_{1} \varepsilon_2}{k_{2} \varepsilon_1} + \frac{k_{2} \varepsilon_1}{k_{1} \varepsilon_2}\right)
sin(k_{1} d_1) sin(k_{2} d_2),
\label{eq_dispersion}
\end{equation}
where $d_i$ and $\varepsilon_i$ (with $i=1,2$) denote the thickness and permittivity of the layers, $a=d_1+d_2$ is the period of the structure, $k_B$ is the Bloch wavevector, $k_i = \sqrt{k_0^2 \varepsilon_i-k_{x}^2}$ is the component of the wavevector along the z-axis inside the i-th medium, and $k_0=2\pi/\lambda$ is the free space wavenumber. A multilayer with $N$ periods,  total thickness $L=N\cdot a$, and with symmetric termination  is shown in Fig.~\ref{fig_layers}. An analogous dispersion relation for a TE-polarised wave was also presented by Wu et.~al.,~\cite{Wu2003} however our interest is focused on metal-dielectric multilayers for the TM polarisation, since they enable  SPP-enhanced transmission with sub-wavelength resolution.

When the layers are thin $k_i\cdot d_i\ll 1$, EMT makes it possible to approximate the  structure with a uniaxially anisotropic slab with the effective permittivity tensor~\cite{wood2006, belov2005csi, belov2006prb}
\begin{equation}
\hat \varepsilon = \vline \begin{array}{ccc} \varepsilon_{x} & 0 & 0 \\ 0 & \varepsilon_{x} & 0 \\ 0 & 0 & \varepsilon_{z}  \end{array} \vline \;,
\label{eq_tensor}
\end{equation}
where
\begin{equation}
\varepsilon_{x} = \frac{\varepsilon_1 d_1 + \varepsilon_2 d_2}{d_1+d_2} \;, \;\;\;\;\;\;\;\;\;
\varepsilon_{z} =\frac{d_1+d_2}{\varepsilon_1^{-1}d_1 + \varepsilon_2^{-1}d_2}
\label{eq_effective}
\end{equation}
are the effective permittivities for the directions parallel and normal to the layers surfaces, respectively. Equations~(\ref{eq_tensor}) and (\ref{eq_effective}) may be derived from the electromagnetic boundary conditions by averaging electric fields {\bf E} and {\bf D} within a single elementary cell of the multilayer.~\cite{markos_wave_prop}

In a lossless periodic layered structure
$\varepsilon_{z}$ may achieve an infinite value if the thicknesses $d_1$, $d_2$, and permittivities  $\varepsilon_1$, $\varepsilon_2$ satisfy the following  relation~\cite{belov2006prb}
\begin{equation}
\frac{\varepsilon_1}{\varepsilon_2}=-\frac{d_1}{d_2}.
\label{eq_canalization}
\end{equation}
In a lossy structure $\varepsilon_{z}$ is always finite but may have a large magnitude which still enables to obtain a device that couples a broad spectrum of spatial frequencies, including both homogeneous and evanescent waves, into a propagating mode. This mode is almost diffraction-free 
 and in low-loss media it may be guided for a large distance. This enables the projection of sub-wavelength images from the front interface of the device onto the back interface and to obtain in-plane imaging with sub-wavelength resolution.

Matching the impedances $\eta=\sqrt{\mu/\varepsilon}$ of two media is a way to avoid reflections from their boundary. In a similar way, assuring in Eq.~(\ref{eq_effective}) that $\varepsilon_x \approx 1$ or
\begin{equation}
\varepsilon_1 d_1 + \varepsilon_2 d_2\approx d_1+d_2
\label{eq_impmatch}
\end{equation}
makes it possible to eliminate reflections from the multilayer for normal incidence. A further increase of transmission and removal of reflections is possible for the Fabry-Perot condition
\begin{equation}
\frac{\sqrt\varepsilon_x \cdot L}{\lambda} \approx \frac{m}{2}, \textrm{ where } m=1,2\ldots
\label{eq_fabryperot}
\end{equation}
Finally, when the size of a single dielectric layer is sufficient to form a cavity in between the metallic layers, cavity modes may be coupled enabling  resonant transmission of a wave through the structure even when the metallic layers are a lot thicker than the skin depth in  bulk metal. The condition for a cavity mode to exist is
\begin{equation}
\frac{\sqrt\varepsilon_1 \cdot d_1}{\lambda}  \approx \frac{l}{2}-\frac{\varphi_{r_{12}}}{2\pi},
\label{eq_cavity}
\end{equation}
where $l=1,2\ldots$ and $\varphi_{r_{12}}$ is the phase of the complex reflection coefficient between materials with permittivities $\varepsilon_1$ and $\varepsilon_2$ responsible for shortening of the cavity length.

In order to measure the transmission efficiency of the multilayer we use the effective skin depth~($\delta$), defined as the distance at which the intensity of a normally incident wave decreases by a factor of the Euler constant $e$. Notably, sometimes the skin depth is defined with respect to amplitude rather than intensity, or the use of this term is restricted to the skin effect due to conductivity of metals, while here we use it in a broader sense~\cite{Saleh} equivalently to the effective intensity decay rate and independently of its physical origin. Therefore the skin depth of a uniform or homogenised medium  may be calculated using the simple formula,
\begin{equation}
\delta = \lambda / {4 \pi \Im{(n_{x})}},
\label{eq_skin-depth}
\end{equation}
where $\Im{(n_{x})}$ is the imaginary part of the effective refractive index in the direction parallel to the multilayer. For non-magnetic materials the effective index may be either defined using the effective permittivity tensor given in Eq.~(\ref{eq_effective}) $n_x\equiv n_x^{EMT} = \sqrt{\varepsilon_x}$  or from the Bloch wavevector calculated using the dispersion relation~(\ref{eq_dispersion}) $n_x\equiv  n_x^{Bloch}=k_B(k_x=0)/k_0$. We underline that the imaginary part of $n_x^{Bloch}$ does not depend on the choice of the Brillouin zone and that the two definitions asymptomatically converge when $a \ll \lambda$.

As the measure of resolution we use the $FWHM$ of the squared point spread function ($|PSF|^2$).~\cite{Kotynski:OL2010} The point spread function is commonly used in Fourier optics to characterize the response of an optical system to a point source, whereas the $|PSF|^2$ is directly related to the intensity distribution in the image plane.

\section{Optimisation results}
\label{results}

We focus on multilayers consisting of silver and three different dielectric or semiconductor materials: TiO$_2$, SrTiO$_3$ and GaP. Their respective dispersion characteristics used in modeling are  taken from Johnson and Christy~\cite{johnson_christy1972} in case of silver and from the book by Palik~\cite{Palik} otherwise. The choice of these materials is based on their large permittivities in the visible range and technical possibility of using them to prepare thin layers with several techniques.~\cite{Boltasseva2010,Duyar_TiO2, Dzibroul_tio2, Das_srtio3_electron_beam, Nam_srtio3_magnetron}

\begin{figure}
 \begin{center}
\includegraphics{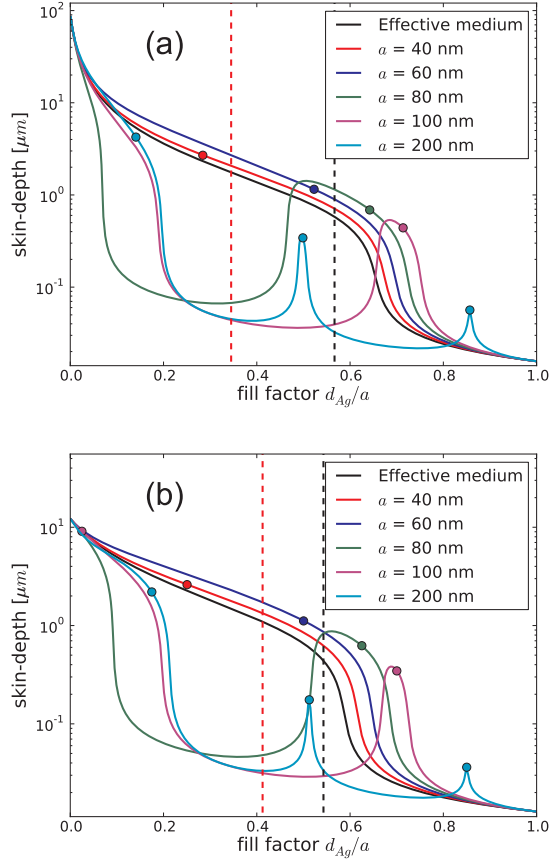}
\end{center}
\caption{Skin depth of a) Ag--TiO$_2$ and b) Ag--GaP periodic multilayer as a function of the fill factor of the structure. Particular curves correspond to different thickness of the elementary cell of the structure, varying from $a = 40$~nm to $a = 200$~nm. The black curve corresponds to the homogenised medium in accordance to the effective medium theory. The wavelength is equal to a) 390~nm and b) 490~nm. The points marked at the curves refer to the values of the fill factor for which the cavity modes  are supported in a single dielectric layer. Dashed lines indicate the filling fractions for diffraction-free propagation (in red) and for impedance matched to air  (in black).}
\label{fig_skin}
\end{figure}

In Fig.~\ref{fig_skin} we present the effective skin depth of Ag--TiO$_2$ and Ag--GaP periodic multilayers as a function of the filling fraction of silver for several values of the period ranging from $a=40$~nm to $a=200$~nm and for the wavelengths of $\lambda = 390$~nm and  $\lambda = 490$~nm, respectively.
 The skin depth  for a periodic structure is calculated with Eq.~(\ref{eq_skin-depth}) using the effective index $n_x\equiv  n_x^{Bloch}$ obtained  from the Bloch wavevector $k_B$ given from the dispersion relation (Eq.~(\ref{eq_dispersion})). Additionally, it is also calculated for an effective medium with $n_x\equiv n_x^{EMT}$, notably the two methods   converge for $a\mapsto 0$. In the opposite limit, for large values of the period $a$, the multilayer behaves in a resonant way remaining transparent only for the size of dielectric layers corresponding to the cavity modes exited in the dielectric layers. The locations of these modes calculated with Eq.~(\ref{eq_cavity}) are shown with circles. In a periodic structure with sufficiently thin metallic layers ($d_{Ag}=a\cdot(d_{Ag}/a)\lesssim40$~nm) these modes broaden into a band, which is reflected by the non-resonant behavior of skin depth in Fig.~\ref{fig_skin} for  small fill-factors. Nonetheless, the specific shape of the field repeated in subsequent coupled cavities is also observed already within the band when the condition~(\ref{eq_cavity}) applies. We will further demonstrate such a case for $Ag-GaP$ and $a=60$.
Moreover, the filling factors which under EMT satisfy the conditions of diffraction-free propagation (Eq.~(\ref{eq_canalization})) and impedance matching (Eq.~(\ref{eq_impmatch})) with air are also marked in the same plot with vertical dashed lines.
Further, we focus on structures with the period of $a=40$~nm and $a=60$~nm, primarily due to the technological feasibility of depositing thick ($L\geqslant\lambda$) multilayers with such a period but also due to their increased skin depth, as compared with the multilayers with a smaller period~(See~Fig.~\ref{fig_skin}).

\begin{figure}
 \begin{center}
\includegraphics{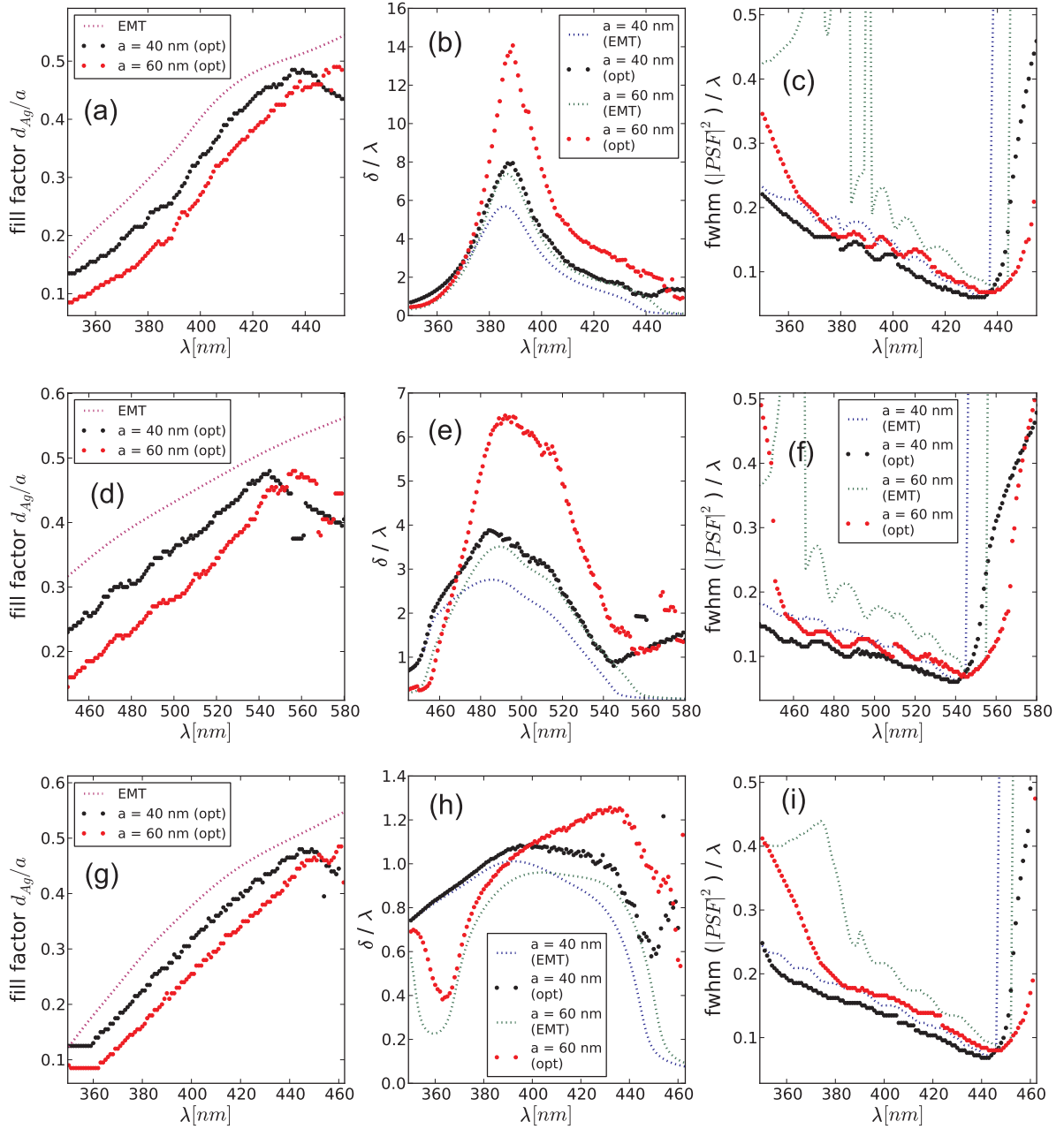}
\end{center}
\caption{Optimization results for Ag--TiO$_2$ (a,b,c), Ag--GaP (d,e,f), and Ag--SrTiO$_3$ (g,h,i) multilayers as a function of wavelength. Sub-figures include the fill factor (a,d,g), effective skin depth $\delta/\lambda$  (b,e,h), and a measure of resolution - $FWHM$ of $|PSF|^2$ (c,f,i). Fill factor is optimised for the sizes of  period  $a=40$~nm and $60$~nm (opt) and is compared to diffraction-free condition (EMT). Resolution is calculated for structures with the thickness of $L=1\lambda$.}
\label{fig_optim}
\end{figure}

Our goal in this paper is to optimize the  resolution and skin depth of multilayers in the visible wavelength range. A periodic multilayer is defined by its total thickness $L$, period $a$, and the fill factor $d_{Ag} / a$. In the optimization procedure the resolution is measured for metal-dielectric stacks of the thickness equal to one wavelength $L=\lambda$ and the fill factor is optimized. The optimisation is of the trade-off character, since the effective skin depth decreases monotonically with the increase of the fill factor $d_{Ag}/a$ while the $FWHM$ of the $|PSF|^2$ takes minima only for specific values of the fill factor.  We use the transfer matrix method (TMM) to calculate the $PSF$. In order to combine  high  resolution and transmission efficiency, we minimise the ratio $FWHM / \delta$ subject to the constraint that $FWHM$ is less than $\lambda / 2$.

\begin{figure}
 \begin{center}
\includegraphics{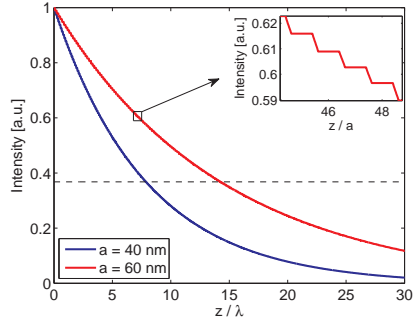}
\end{center}
\caption{1D FDTD simulation with time-averaged Poynting vector inside the Ag--TiO$_2$ structure for $a=40$~nm  (lower blue line) and  for $a = 60$~nm (upper red line) at the wavelength of $\lambda = 390$~nm. The fill factors of both structures are optimised. The intensity level of $1/e$ is marked with a dashed line. The inset shows a magnified part of the plot. }
\label{fig_fdtd_1d_Sz}
\end{figure}

In Fig.~\ref{fig_optim} we present the optimization results for Ag--TiO$_2$, Ag--SrTiO$_3$ and Ag--GaP structures, respectively.
The resolution and the effective skin depth of the optimised multilayers are compared with those with the fill factor obtained using the EMT-based expression for diffraction-free propagation, Eq.~(\ref{eq_canalization}). Clearly, for a broad range of wavelengths, optimization leads to the improvement of both resolution and transmission efficiency at the same time. 
The largest skin depth is achieved for Ag--TiO$_2$ structure for $\lambda = 390$~nm, and for Ag--GaP structure for $\lambda = 490$~nm.
They are equal to $\delta = 14~\lambda$ and to $\delta = 6.5~\lambda$, respectively. In both cases the resolution is on the order of $0.12-0.15~\lambda$ and it is little affected by a change in $a$. For Ag--SrTiO$_3$ the skin depth does not exceed $1.3~\lambda$ for any wavelength, which makes strontium titanate less interesting for fabricating thick multilayers, although the maximal value of skin depth corresponds to the resolution better than $0.1~\lambda$.
 The skin depth is larger for structures with the larger period $a = 60$~nm than for those with $a = 40$~nm. We illustrate this increase with a one-dimensional FDTD simulation for the Ag--TiO$_2$  stack shown in Fig.~\ref{fig_fdtd_1d_Sz}. The plot includes the time-averaged Poynting vector inside an infinite structure. For both periods, light decays exponentially, however the larger thickness of elementary cell results in a slower decay rate. The inset in Fig.~\ref{fig_fdtd_1d_Sz} contains a magnified  part of the same plot showing that the energy is absorbed only in the silver layers, since titanium dioxide is practically lossless in the visible range. The decay rate obtained with FDTD is in perfect agreement with the value of effective skin depths calculated with the dispersion relation (\ref{eq_dispersion}).

\begin{figure}
 \begin{center}
\includegraphics{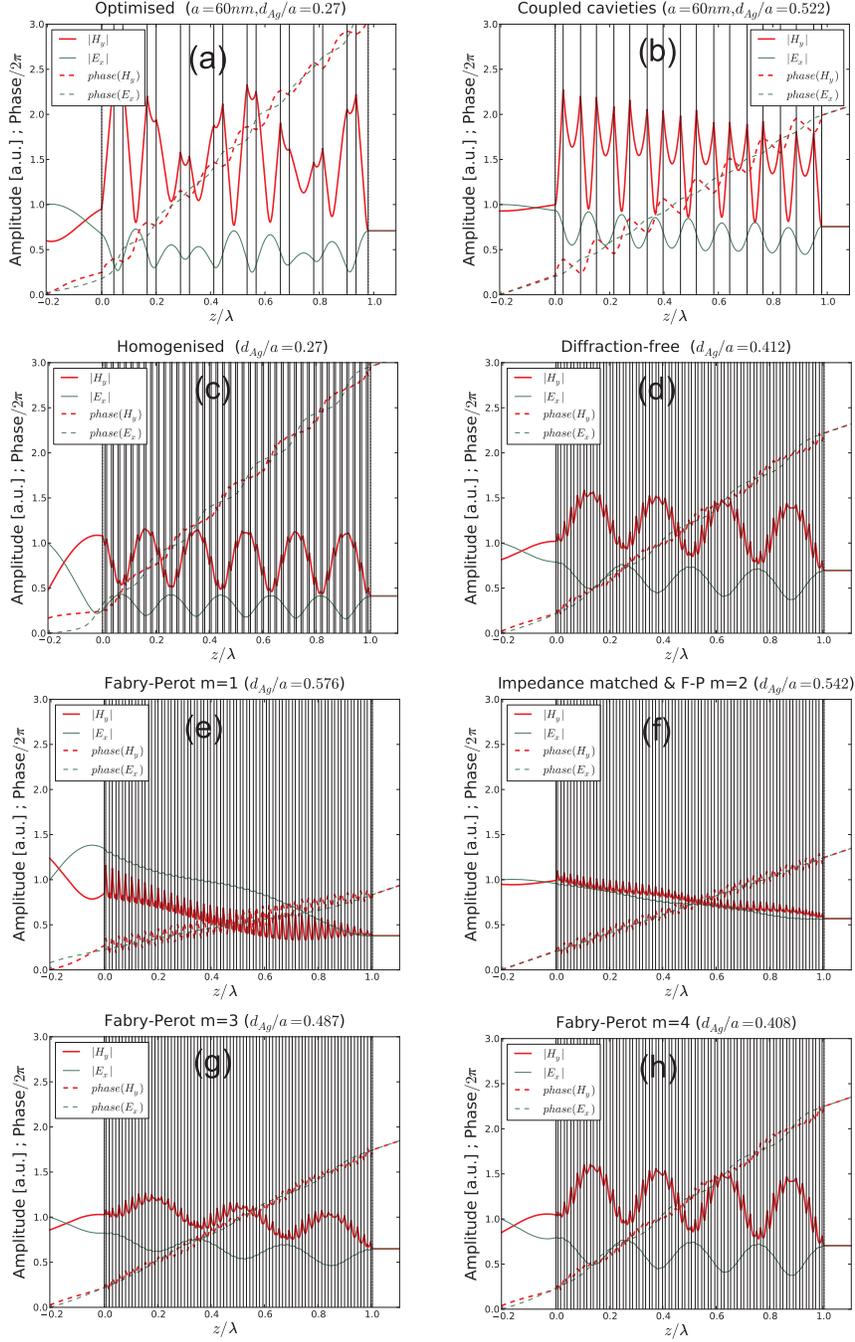}
\end{center}
\caption{Transmission of a normally incident plane wave through  Ag--GaP multilayers with the total thickness $L=\lambda$ at the wavelength of $\lambda=490$~nm. a) multilayer optimised for $a=60$~nm; b) coupled-cavity multilayer with  $a=60$~nm; c) homogenised version of (a); d) multilayer with maximised $\varepsilon_z$;  e-h) multilayers satisfying the Fabry-Perot condition of the order $m=1,2,3,4$. Multilayer with $m=2$ is at the same time impedance matched to air. }
\label{fig_field}
\end{figure}

\begin{table}
\caption{Comparison of the effective skin-depths~($\delta$), resolutions~(${FWHM}(|PSF|^2)$) and intensity transmission coefficients for normal incidence ($T$) of the Ag--GaP multilayers with the total thickness $L=\lambda$ at the wavelength of $\lambda=490$~nm.}
\label{tab.resolution}
\begin{tabular}
{  p{6cm}  l l l l l}
\hline
Description of the multilayer& $d_{Ag}/a$ & $\delta/\lambda$ & $\frac{FWHM}{\lambda}$ & $T$  \\
\hline
\hline
Optimised for $a=60$ nm & $0.270$ & 6.31 & 0.12 & 0.78\\
Coupled cavities with $a=60$ nm & $0.522$ & 2.00 & 0.58 & 0.61\\
\hline
Optimised and homogenised & $0.270$ & 4.22 & 0.76 & 0.42\\
Diffraction-free & $0.412$ & 2.23 & 0.08 & 0.58\\
Fabry-Perot $m=1$ & $0.576$ & 0.45 & 0.08 & 0.10\\
Impedance-matched, F-P $m=2$ & $0.542$ & 0.90 & 0.11 & 0.33\\
Fabry-Perot $m=3$ & $0.487$ & 1.47 & 0.11 & 0.49\\
Fabry-Perot $m=4$ & $0.408$ & 2.27 & 0.08 & 0.59\\
\hline
\end{tabular}
\end{table}

Good insight into the transmission mechanism involved in the imaging through the multilayer is gained from the internal field distribution. With this aim, in Fig.~\ref{fig_field} we summarise the field profiles for several designs of  Ag--GaP structure at $\lambda=490$~nm with the total thickness of $L=1\lambda$, either with $a=60$~nm or with extremely thin layers $a\mapsto 0$. We compare the following structures: the one which we have optimised, the one consisting of coupled-cavities (with the width of $d_2$ defined using Eq.~({\ref{eq_cavity}) for the fundamental mode), the homogenised analogue of the optimised one (split into more layers with the filling fraction and total thickness fixed), a diffraction-free structure (fill factor calculated with Eq.~(\ref{eq_canalization})), and Fabry-Perot etalons of the order $m=1,2,3,4$ (with $m=2$ also satisfying the condition for impedance matching~(\ref{eq_impmatch})). The corresponding filling fractions, skin depths, resolutions and intensity transmission coefficients are compared in Tab.~\ref{tab.resolution}. The immediate conclusion from this summary is that the optimised multilayer does not resemble the theoretical designs neither in terms of the filling fraction nor in terms of the internal field distribution. This is a strong argument for using numerical optimisation in place of simplified models based on EMT when it is important to combine a good transmission efficiency, a subwavelength resolution and to keep the total number of layers technologically feasible.

In Fig.~\ref{fig_TF_phase} we present the amplitude and phase of amplitude transfer functions (TF) for Ag--TiO$_2$ multilayer (at $\lambda = 390$~nm) and for Ag--GaP multilayer (at $\lambda = 490$~nm). The period $a$ and the total thickness of structure $L$ are fixed at $60$~nm and $3~\lambda$, respectively. Horizontal cross-sections of  Fig.~\ref{fig_TF_phase} include the TF  calculated for a range of fill factors, and these TF depend on $k_x/k_0$.

For propagating waves $k_x / k_0$ is the sine of the angle of incidence. The values of wavevector $k_x / k_0 > 1$ refer to the evanescent waves in air. Optimal fill factors for both structures are $0.22$ and $0.28$ for Ag--TiO$_2$ and Ag--GaP multilayers, respectively. At the same time, for these values of the fill factor, the  phase of TF in Fig.~\ref{fig_TF_phase} is the flattest, while the amplitude is slowly varying. The phase is practically independent of $k_x / k_0$ and is similar for propagating and for evanescent components of the spatial spectrum enabling diffraction-free propagation with sub-wavelength resolution. 
Moreover, sub-wavelength resolution may be obtained even if the overall thickness of the multilayer is much larger than the wavelength.

\begin{figure}
 \begin{center}
\includegraphics{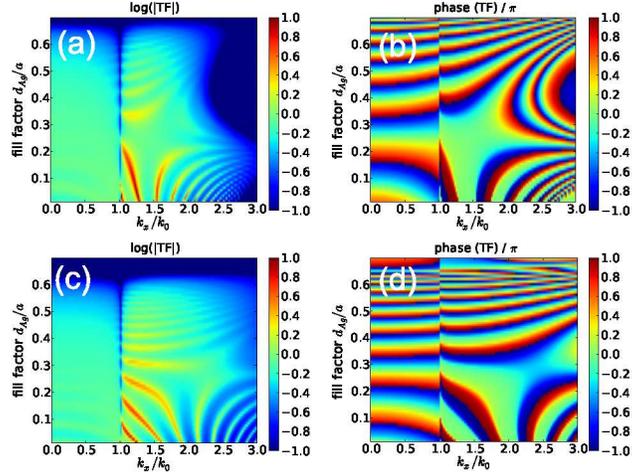}
\end{center}
\caption{Dependence of the amplitude transfer function TF (in horizontal cross-sections of the figure) on the fill factor: a, c) logarithm of amplitude, and  b, d) phase. Two structures were considered: a, b) Ag--TiO$_2$ multilayer (for $\lambda = 390$~nm) and c, d) Ag--GaP multilayer (for $\lambda = 490$~nm). The total thickness of both structures is equal to $3~\lambda$ and the thickness of the elementary cell $a = 60$~nm. The optimal fill factors  are $0.22$ for Ag--TiO$_2$  and $0.28$ for Ag--GaP.
}
\label{fig_TF_phase}
\end{figure}

\begin{figure}
 \begin{center}
\includegraphics{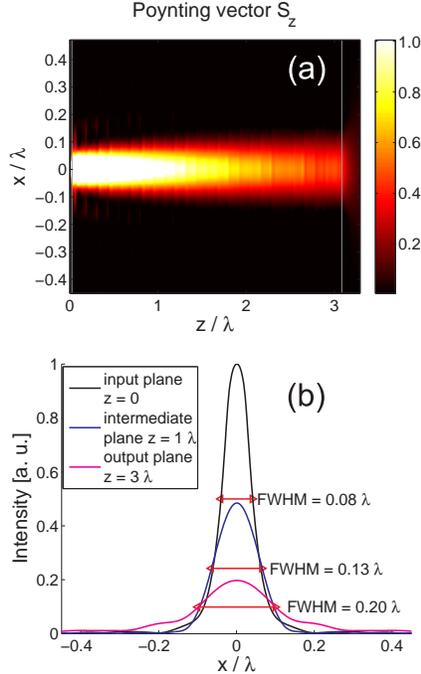}
\end{center}
\caption{FDTD simulation of Ag--GaP multilayer consisting of 25 elementary cells with $a = 60$~nm and fill factor 0.28. The structure is illuminated with a plane wave with $\lambda = 490$~nm diffracted at a subwavelength aperture with the diameter equal to $\lambda /20$. a)~The z-component $S_z$ of Poynting vector; b)~intensity profiles in three cross-sections: at input and output interfaces of the multilayer and at  $z=1 \lambda$ from the input interface. }
\label{fig_FDTD}
\end{figure}
Let us now illustrate the performance of the optimised multilayer. Up to now we have neglected any possible interaction between the source of the sub-wavelength-sized wavefront and the multilayer. In practice, the source may be realised using a metallic mask with sub-wavelength aperture, a kinoform with sub-wavelength relief, a plasmonic waveguide coupled to the multilayer, or with a SNOM probe. In any case, the source needs to be put in the near field or preferably be attached to the multilayer, therefore causing multiple reflections in between both elements. The system consisting of the source coupled to the multilayer analyzed as a whole, may be no longer shift invariant, although  idealized source models such as a soft (uncoupled) source, or hard (strongly coupled) magnetic and electric sources may still be included in the same framework of LSI~\cite{Kotynski:oer2010}. The three source models become equivalent in the absence of reflections~\cite{Kotynski:oer2010}. Even though, a soft source model is assumed by us in this paper, in case of the optimised multilayer,  thanks to small reflections, the FWHM of PSF is nearly the same for the three source models. 

In Fig.~\ref{fig_FDTD} we present the results of a two-dimensional FDTD simulation of a Ag--GaP structure consisting of $25$ elementary cells with $a = 60$~nm and fill factor equal to $d_{Ag} / a = 0.28$. We show the  longitudinal component of the Poynting vector $S_z$ and intensity profiles at three cross-sections perpendicular to the propagation direction. The structure is illuminated with a monochromatic planewave with $\lambda = 490$~nm diffracted at a sub-wavelength aperture with the diameter equal to $\lambda /20$. The aperture is made in a perfect metal screen having the thickness of $2$~nm located at the distance of $10$~nm from the multilayer. Transmission through the aperture is not enhanced by resonant effects. Resonant transmission would certainly complicate our analysis, whilst our main focus here is on the transmission properties of the multilayer rather than on the efficient wavefront modulation at the scale of $20$~nm. This way, the presented results are general in the sense that they are little affected by the choice of metal used to make the mask, provided that its thickness is sufficient to make it opaque. For instance, a chromium mask with the thickness of $50$~nm could be used instead the perfect conductor giving similar results.

The sub-wavelength dimension and shape of the wavefront is preserved within the structure despite the continuous intensity decay due to the  losses in metallic layers. Moreover, the $FWHM$ at the distance of $1 \lambda$ is consistent with the optimisation results obtained with the transfer matrix method and presented in Fig.~\ref{fig_optim}f.

\section{Conclusions}
\label{conclus}

We have optimised the effective skin-depth and resolution of Ag--TiO$_2$, Ag--SrTiO$_3$, and Ag--GaP multilayers for imaging with sub-wavelength resolution.
We have shown that  multilayers with a period of $a=40$~nm ($\approx8\%\cdot\lambda$) or larger,  designed using the effective medium theory (Eqs.~(\ref{eq_canalization}),(\ref{eq_impmatch}),(\ref{eq_fabryperot})) are suboptimal in terms of both resolution and effective skin depth at the same time. Still, multilayers with a large thickness of layers and with silver layers slightly thicker than the skin depth may combine a good trade-off between resolution and transmission. Optimal multilayer designs with a period of $a=40$~nm or $a=60$~nm outperform multilayers based on combined use of effective medium theory~(with $a/\lambda\mapsto 0$, $N=L/a$ and $L=const$), impedance matching and Fabry-Perot resonances  in terms of transmission, with a certain deterioration of resolution.
 For instance, an optimised Ag--GaP multilayer consisting of only $17$ layers, operating at the wavelength of $490$~nm and having the total thickness equal to the wavelength, combines the intensity transmission coefficient of $78\%$ with the resolution of $60$~nm. Using EMT, the resolution may be improved to $37$~nm (See Tab.~\ref{tab.resolution}) but to achieve this it is necessary to increase the number of layers several-fold which is technologically challenging. We have shown that an optimised structure allows for diffraction-free guidance of a sub-wavelength sized beam of light for at least $L=3\cdot \lambda$ and that the transmission is not directly based on Fabry-Perot resonances. At the same time, the optimal design does not resemble the EMT-based multilayer nor the coupled-cavity multilayer in terms of either the filling fraction or the internal distribution of field making an argument for the use of numerical optimisation instead of relaying on simplified theoretical designs.

\section{Acknowledgements}
We acknowledge support from the Polish Ministry of Science and Higher Education research project N N202 033237, the
(Polish) National Centre for Research and Development research project N R15 0018 06, and the framework of European Cooperation
in Science and Technology -- COST actions MP0702, MP0803.

%

\end{document}